%% file: lightgcn_extensions_article_with_mlp_full.tex
\documentclass[10pt,twocolumn]{article}
\usepackage[margin=0.78in]{geometry}
\usepackage{amsmath,amssymb,booktabs,graphicx,hyperref}
\usepackage[T1]{fontenc}
\usepackage{lmodern}
\usepackage{microtype}
\title{Sheaf-Inspired Structured Scoring Extensions of LightGCN for Collaborative Filtering}
\author{Ksenia Belkovskaya \and
    Timur Lidzhiev \and
    Maria Stebliy \and
    Alexander Tarakanov}
\date{}
\begin{document}
\maketitle
\begin{abstract}
Graph-based collaborative filtering learns user and item representations by propagating embeddings over a user--item interaction graph. We investigate a modular, sheaf-inspired scoring extension placed on top of a three-layer LightGCN backbone. The motivation comes from cellular sheaves, where local vector spaces are related by linear restriction maps; here, this idea motivates structured, pair-dependent transformations of user and item embeddings rather than a change to graph propagation. We compare the identity inner-product scorer, diagonal, rank-one low-rank, and full matrix residual scorers, and a parameter-matched residual MLP baseline (\texttt{mlp\_full}) on Amazon-Book, Gowalla, MovieLens-1M, and Yelp2018. We report ranking quality and catalog coverage at cutoffs 10 and 20. The selected structured variants show modest, dataset-dependent changes in ranking quality and sometimes substantial changes in coverage. Our method is inspired by sheaf geometry but does not construct a cellular sheaf or implement sheaf-Laplacian diffusion; the experiments therefore do not establish a causal benefit from sheaf geometry itself.
\end{abstract}

\section{Introduction}
Graph-based collaborative filtering has become an important
approach to modern recommender systems, as user--item
interactions can be naturally represented by a bipartite
graph. LightGCN~\cite{he2020lightgcn} demonstrated that
simplifying graph convolution by removing feature
transformations and nonlinear activations can improve
recommendation performance while reducing model complexity.

Despite its effectiveness, conventional LightGCN relies
on the inner product of learned user and item embeddings
to estimate their compatibility. This scoring mechanism
implicitly assumes a shared embedding coordinate system
and does not explicitly model heterogeneous transformations
between user and item representations.

Cellular sheaf theory provides a mathematical framework
for describing locally varying relationships through
vector spaces and linear restriction maps
~\cite{curry2014sheaves,hansen2019spectral}.
Building on this framework, Sheaf Neural Networks
~\cite{hansen2020sheaf} and Neural Sheaf Diffusion
~\cite{bodnar2022neural} demonstrated how learned
geometric structures can generalize conventional
graph diffusion and address challenges such as
heterophily and oversmoothing.

More recently, sheaf-based methods have been
investigated for recommendation tasks.
Sheaf4Rec~\cite{purificato2025sheaf4rec} demonstrated
the applicability of sheaf neural networks to
graph-based collaborative filtering. However,
introducing sheaf-based message passing may increase
model complexity compared with standard LightGCN.

The central idea of this work is to place a lightweight,
\emph{sheaf-inspired geometric scoring layer on top of LightGCN}.
The LightGCN backbone first produces user and item embeddings
using its original message-passing rule. An additional
scoring module then learns a pair-dependent linear
transformation between these representations, motivated
by the restriction maps that relate local vector spaces
in cellular sheaves~\cite{curry2014sheaves,hansen2019spectral}.
Thus, the geometric component is an optional extension
of the compatibility function, not a replacement for
the graph encoder. This modular design lets us test
whether imposing structure on the score correction
provides a useful inductive bias beyond a generic
neural-network correction.

Importantly, we use \emph{sheaf-inspired} in a specific,
limited sense: we borrow the idea of structured maps
between representation spaces, but do not assign
formal stalks and restriction maps to every vertex
and edge or enforce sheaf consistency. In particular,
our model does not introduce sheaf-Laplacian diffusion
into LightGCN propagation.

We evaluate several transformation families,
including diagonal, full, and low-rank operators,
and compare them against the standard inner-product
scorer and a parameter-matched residual MLP baseline.

Our experiments are conducted on four recommendation
datasets: Amazon-Book, Gowalla, MovieLens-1M,
and Yelp2018. We assess ranking quality using
Recall, NDCG, MRR, and Precision, together with
catalog coverage at cutoffs 10 and 20.

\section{Related Work}

\subsection{Mathematical Foundations of Cellular Sheaves}

Cellular sheaves provide a general framework for
representing local-to-global consistency in
structured spaces. Curry~\cite{curry2014sheaves}
established mathematical foundations for sheaves
and cosheaves, including their applications to
discrete structures.

Hansen and Ghrist~\cite{hansen2019spectral}
developed a spectral theory of cellular sheaves,
introducing sheaf Laplacians that generalize
classical graph Laplacians. Unlike conventional
graph operators, sheaf Laplacians incorporate
linear restriction maps connecting vector spaces
associated with graph elements.

These constructions provide a theoretical
motivation for structured transformations
between local representations.

\subsection{Sheaf Neural Networks}

Hansen and Gebhart~\cite{hansen2020sheaf}
introduced Sheaf Neural Networks, extending
graph neural network architectures through
sheaf-based diffusion operators.

Bodnar et al.~\cite{bodnar2022neural}
proposed Neural Sheaf Diffusion, demonstrating
that learnable sheaf structures can improve
representation learning on graphs with
heterophilous relationships and mitigate
oversmoothing.

Barbero et al.~\cite{barbero2022sheaf}
investigated Sheaf Neural Networks with
Connection Laplacians, exploring structured
geometric operators for graph learning.

Subsequent studies extended sheaf-based
representation learning in several directions.
Duta et al.~\cite{duta2023sheaf}
generalized sheaf neural networks to hypergraphs,
allowing higher-order interactions to be modeled.

He et al.~\cite{he2024sheaf}
introduced sheaf-based positional encodings,
demonstrating how sheaf operators can provide
structural information for graph neural networks.

Zaghen et al.~\cite{zaghen2024sheaf}
investigated nonlinear diffusion using adaptive
sheaf Laplacians, while Hernandez Caralt
et al.~\cite{hernandez2024joint}
studied joint diffusion processes as an
inductive bias in sheaf neural networks.

Together, these studies highlight the importance
of structured geometric operators for learning
expressive graph representations.

\subsection{Graph-Based Recommendation}

LightGCN~\cite{he2020lightgcn} established
a simplified graph-based collaborative filtering
architecture by eliminating unnecessary feature
transformations and nonlinear activation functions
from graph convolution.

Self-supervised Graph Learning for Recommendation
(SGL)~\cite{wu2021self} further demonstrated
the effectiveness of contrastive learning
for improving graph-based recommendation
representations.

More recently, Purificato et
al.~\cite{purificato2025sheaf4rec}
proposed Sheaf4Rec, adapting sheaf neural
networks to graph-based recommendation tasks.

Unlike methods that introduce sheaf operators
into message passing, our approach retains
standard LightGCN propagation and applies
lightweight structured transformations
during user--item scoring.

This design allows us to investigate whether
geometrically motivated parameterizations
offer advantages over both conventional
inner-product scoring and unstructured
neural scoring functions.

\section{Model and Training}
\subsection{LightGCN backbone}
Let $\mathcal{U}$ and $\mathcal{I}$ be the user and item sets and let $A$ represent the undirected bipartite training graph. Its symmetrically normalized adjacency is $\widetilde A=D^{-1/2}AD^{-1/2}$. The initial trainable embeddings are stacked into $E^{(0)}$. For layers $\ell=1,\ldots,L$, propagation is
\begin{equation}
E^{(\ell)}=\widetilde A E^{(\ell-1)},\qquad
Z=\frac{1}{L+1}\sum_{\ell=0}^{L} E^{(\ell)}.
\end{equation}
The score for user $u$ and item $i$ is $\hat y_{ui}=z_u^\top z_i$. The inspected code uses a mean over all layer embeddings, including layer zero, with $L=3$ and embedding dimension $d=64$.

Training uses Bayesian personalized ranking (BPR) \cite{rendle2009bpr}. For a sampled positive interaction $(u,i)$ and negative item $j$, the ranking term is $\operatorname{softplus}(\hat y_{uj}-\hat y_{ui})$. The loss adds regularization over the \emph{initial} embeddings of the sampled user, positive item, and negative item:
\begin{equation}
\mathcal L=\mathcal L_{\mathrm{BPR}}+\lambda_{2}\mathcal L_{2},\quad
\mathcal L_{2}=\frac{\|e_u^{(0)}\|_2^2+\|e_i^{(0)}\|_2^2+\|e_j^{(0)}\|_2^2}{2B},
\end{equation}
where $B$ is the minibatch size and the numerator is summed over its triples.

\subsection{Structured scoring extensions and MLP baseline}
We interpret the extension as a geometrically motivated \emph{post-encoder layer}: LightGCN generates embeddings, and the scorer applies pair-dependent linear maps to those embeddings. This architectural separation is intentional. Cellular sheaves motivate the use of maps between local vector spaces~\cite{hansen2019spectral,hansen2020sheaf}, but the maps below are learned scoring corrections, not formal sheaf restriction maps. The supplied \texttt{sheafrec} code applies these post-encoder scoring functions to user and item embeddings $u,v\in\mathbb{R}^{64}$. Its input features are
\begin{equation}
\phi(u,v)=[u;v;u\odot v;|u-v|].
\end{equation}
The identity reference uses $s_{\mathrm{id}}(u,v)=u^\top v$. A structured scorer produces a pair-dependent matrix $\Delta_m(u,v)$ and computes
\begin{equation}
 s_m(u,v)=u^\top v+\alpha u^\top\Delta_m(u,v)v.
\end{equation}
For \texttt{diagonal}, $\Delta_m$ is diagonal; for \texttt{low\_rank\_r1}, it factors as a rank-one matrix; and for \texttt{full}, the predicted residual is a dense matrix. We treat these parameterizations as progressively more flexible choices for a sheaf-inspired geometric layer. Their role is analogous to a learned transformation between the user and item embedding spaces, but they are not sheaf restriction maps in the formal mathematical sense. The predicted matrices do not modify the LightGCN propagation matrix, and we do not build a sheaf Laplacian.

To control for the possibility that gains merely arise from adding a neural scorer, we include \texttt{mlp\_full}. It uses the same pair features and a residual dot-product skip:
\begin{equation}
 s_{\mathrm{MLP}}(u,v)=u^\top v+\alpha f_{\theta}(\phi(u,v)).
\end{equation}
The implementation uses two hidden linear layers with GELU and dropout, followed by a scalar output; their widths are chosen to approximately match the parameter budget of the \texttt{full} structured scorer. Matching parameter counts is a useful control but does not equalize all inductive biases or computational costs. The supplied aggregate results cover multiple implementations; full experimental provenance and hyperparameter parity should be verified against run logs before submission.

\subsection{Separate contrastive-learning implementation}
The provided CL code constructs two independent views by dropping training interactions with probability 0.1 per view, recalculates graph normalization, and computes symmetric InfoNCE between embeddings of matching user or item IDs. It uses temperature $\tau=0.2$, up to 256 users and 256 items per minibatch, and an objective
\begin{equation}
\mathcal L_{\mathrm{CL\ run}}=\mathcal L_{\mathrm{BPR}}+\lambda_2\mathcal L_2+0.05(\mathcal L_{\mathrm{user}}^{\mathrm{NCE}}+\mathcal L_{\mathrm{item}}^{\mathrm{NCE}}).
\end{equation}
The CL experiment is \emph{not} represented by the identity-versus-extension results reported below, so no performance conclusion about CL is drawn from these tables.

\section{Experimental Setup}
\paragraph{Datasets and evaluation.} We analyze Amazon-Book, Gowalla, MovieLens-1M (ML-1M), and Yelp2018 using the supplied aggregated results for the \texttt{test} split only. The output reports Recall, normalized discounted cumulative gain (NDCG), mean reciprocal rank (MRR), Precision, and item Coverage, each at $K\in\{10,20\}$. Higher values are better for all five measures. Coverage reflects the variety of items appearing in recommendation lists; its exact denominator and evaluation population should be confirmed from the original experiment's evaluation code.

\paragraph{Optimization protocol.} The available LightGCN implementation and project README specify Xavier initialization, Adam ($10^{-3}$ learning rate), explicit embedding regularization ($10^{-4}$), batch size 1024 (2048 for Amazon-Book), at most 1000 epochs, and checkpoint selection by validation Recall@20 with patience of 50 epochs. It uses saved data splits, excludes known interactions from ranking candidates, and evaluates the full item ranking. These are documented properties of the provided LightGCN/CL code, but should be cross-checked against the actual extension experiment logs before being stated as settings of every variant.

\paragraph{Seeds and comparison rule.} The Gowalla, ML-1M, and Yelp2018 test results aggregate five seeds (0--4); Amazon-Book test results contain one seed (0). For readability, we show means without standard deviations; thus the tables do not establish statistical significance. For each dataset, the displayed structured comparator was chosen by its \emph{test} NDCG@20 among the supplied variants. This is a descriptive post-hoc selection, not a valid held-out model-selection protocol. A final study should choose each comparator on the validation set and then report its test performance without reselection.

\section{Results}
\input{test_identity_structured_mlp_full.tex}
\paragraph{Ranking quality.} The structured variants and the parameter-matched MLP yield relatively small differences on most ranking metrics. At $K=20$, diagonal improves NDCG on Amazon-Book from 0.031028 to 0.031302, rank-one low-rank improves it on Gowalla from 0.143631 to 0.143908, and full improves it on ML-1M from 0.053299 to 0.054768. On Yelp2018, diagonal changes NDCG@20 from 0.050433 to 0.050475 while the identity reference retains slightly higher MRR@20. At $K=10$, identity is marginally better than diagonal across all four reported accuracy measures on Yelp2018.

\paragraph{Structured models versus the MLP control.}
At $K=20$, the selected structured model exceeds \texttt{mlp\_full} in NDCG on Amazon-Book (0.031302 vs. 0.030826), Gowalla (0.143908 vs. 0.141496), ML-1M (0.054768 vs. 0.054581), and Yelp2018 (0.050475 vs. 0.049711). However, the MLP often achieves greater catalog coverage: Coverage@20 is 0.621830 on Amazon-Book, 0.615427 on Gowalla, 0.728274 on ML-1M, and 0.393734 on Yelp2018. Hence the structured scorers are not unconditionally superior; there is a ranking--coverage trade-off. Moreover, these comparisons are descriptive, and the structured candidates were chosen post hoc from test results.

\section{Limitations and Future Work}
Several limitations constrain interpretation. First, although the structured scoring architectures are documented in the new code archive, complete correspondence between the reported run logs and the supplied source should be verified. Second, model comparators were selected using test NDCG@20, introducing selection bias. Third, Amazon-Book contains only one test seed, and no inferential test is presented for the five-seed datasets. Fourth, matching an MLP to the full scorer by parameter count does not equalize training or inference compute, and the selected diagonal or rank-one models may have different budgets. Before publication, we plan to verify identical data splits and training protocols, select models solely by validation performance, and report paired seed-level differences and runtime/memory costs. No current experiment isolates sheaf-geometric constraints or sheaf diffusion, so any claim that sheaf geometry itself caused improvements would be premature.

\section{Conclusion}
We studied a modular way to place sheaf-inspired, structured scoring transformations on top of a LightGCN backbone without changing its message passing. Across four datasets, the supplied results suggest modest and dataset-dependent NDCG@20 gains from selected structured extensions over identity and the matched MLP, alongside potentially larger differences in coverage. These results support further investigation of structured geometric inductive biases in scoring; they do not establish the effect of a formally defined sheaf or sheaf diffusion. Future work should validate comparator selection, report seed-level uncertainty, and directly test explicit sheaf constraints.

\end{document}

%% file: test_identity_structured_mlp_full.tex
% Test-only comparative results; all entries are mean metrics, larger is better.
% Selected structured extensions were identified post hoc by test NDCG@20.
% Requires booktabs.

\begin{table*}[t]
\centering
\small
\setlength{\tabcolsep}{4.4pt}
\begin{tabular}{llccccc}
\toprule
Dataset & Model & Recall@10 & NDCG@10 & MRR@10 & Precision@10 & Coverage@10 \\
\midrule
Amazon-Book & identity & 0.023226 & 0.024627 & 0.054500 & 0.018927 & 0.323999 \\
 & diagonal & \textbf{0.023516} & \textbf{0.024974} & \textbf{0.055128} & \textbf{0.019180} & 0.354141 \\
 & mlp\_full & 0.023083 & 0.024355 & 0.053211 & 0.018703 & \textbf{0.479186} \\
\midrule
Gowalla & identity & 0.119028 & 0.129695 & 0.261143 & 0.072200 & 0.365033 \\
 & low\_rank\_r1 & \textbf{0.119319} & \textbf{0.129984} & \textbf{0.261937} & \textbf{0.072261} & 0.359962 \\
 & mlp\_full & 0.117079 & 0.127674 & 0.257245 & 0.071279 & \textbf{0.411322} \\
\midrule
ML-1M & identity & 0.080007 & 0.040098 & 0.028146 & 0.008001 & 0.497867 \\
 & full & \textbf{0.081696} & 0.040953 & 0.028738 & \textbf{0.008170} & 0.506724 \\
 & mlp\_full & 0.081000 & \textbf{0.040983} & \textbf{0.028949} & 0.008100 & \textbf{0.598866} \\
\midrule
Yelp2018 & identity & \textbf{0.035973} & \textbf{0.041142} & \textbf{0.096698} & \textbf{0.032016} & 0.209488 \\
 & diagonal & 0.035946 & 0.041121 & 0.096627 & 0.032014 & 0.230798 \\
 & mlp\_full & 0.035528 & 0.040631 & 0.095611 & 0.031650 & \textbf{0.277013} \\
\bottomrule
\end{tabular}
\caption{Test results at $K=10$: identity, the selected structured transport-inspired extension, and a parameter-matched residual MLP (\texttt{mlp\_full}). Bold indicates the best mean across the three methods for each dataset and metric.}
\label{tab:test_threeway_k10}
\end{table*}

\begin{table*}[t]
\centering
\small
\setlength{\tabcolsep}{4.4pt}
\begin{tabular}{llccccc}
\toprule
Dataset & Model & Recall@20 & NDCG@20 & MRR@20 & Precision@20 & Coverage@20 \\
\midrule
Amazon-Book & identity & 0.039582 & 0.031028 & 0.060259 & 0.016422 & 0.456501 \\
 & diagonal & \textbf{0.039775} & \textbf{0.031302} & \textbf{0.060803} & \textbf{0.016525} & 0.484296 \\
 & mlp\_full & 0.039476 & 0.030826 & 0.058938 & 0.016318 & \textbf{0.621830} \\
\midrule
Gowalla & identity & 0.168863 & 0.143631 & 0.268220 & 0.051729 & 0.566477 \\
 & low\_rank\_r1 & \textbf{0.169097} & \textbf{0.143908} & \textbf{0.269001} & \textbf{0.051761} & 0.559728 \\
 & mlp\_full & 0.166579 & 0.141496 & 0.264345 & 0.051142 & \textbf{0.615427} \\
\midrule
ML-1M & identity & 0.132781 & 0.053299 & 0.031694 & 0.006639 & 0.605077 \\
 & full & \textbf{0.136889} & \textbf{0.054768} & 0.032456 & \textbf{0.006844} & 0.609560 \\
 & mlp\_full & 0.135299 & 0.054581 & \textbf{0.032613} & 0.006765 & \textbf{0.728274} \\
\midrule
Yelp2018 & identity & 0.061375 & 0.050433 & \textbf{0.105331} & 0.027539 & 0.306823 \\
 & diagonal & \textbf{0.061509} & \textbf{0.050475} & 0.105303 & \textbf{0.027576} & 0.327045 \\
 & mlp\_full & 0.060481 & 0.049711 & 0.104117 & 0.027140 & \textbf{0.393734} \\
\bottomrule
\end{tabular}
\caption{Test results at $K=20$: identity, the selected structured transport-inspired extension, and a parameter-matched residual MLP (\texttt{mlp\_full}). Bold indicates the best mean across the three methods for each dataset and metric.}
\label{tab:test_threeway_k20}
\end{table*}